\documentstyle[aps]{revtex}
\begin{document}
\tightenlines
\title{Warm Inflation: Towards a realistic
COBE data power spectrum for matter and metric thermal
coupled fluctuations}
\author{Mauricio Bellini\footnote{E-mail: mbellini@mdp.edu.ar}}
\address{Departamento de F\'{\i}sica, Facultad de Ciencias Exactas
y Naturales, \\
Universidad Nacional de Mar del Plata, De\'an Funes, (7600)
Mar del Plata, Buenos Aires, Argentina}
\maketitle
\begin{abstract}
I consider the COBE data coarse - grained
field that characterize the now observable universe for a model
of warm inflation which takes into account the
thermal coupled fluctuations
of the scalar field with the thermal bath.
The power spectrum for both, matter and metric fluctuations
are analyzed.
I find that the
amplitude for the fluctuations of the metric
when the horizon entry,
should be very small for the expected values of temperature.
\end{abstract}
\vskip 3cm
\noindent
PACS number(s): 98.80.Cq.; 04.62.+v
\twocolumn
\section{Introduction}
Warm inflation takes into account separately, the matter and radiation
energy fluctuations. In this scenario the matter field $\varphi$
interacts with the particles of a thermal bath with mean temperature
$T_r$, which is smaller than the Grand Unified Theories (GUT) critical
temperature $T_r < T_{GUT} \simeq 10^{15}$ GeV. This scenario was
firstly studied by Berera\cite{1}. The warm inflation scenario
served as
a explicit demonstration that inflation can occur in the presence
of a thermal component.
In the formalism developed by Berera the temperature of the universe
remains constant during the inflationary expansion.

Warm inflation was originally formulated in a phenomenological setting,
but some attemps of a
fundamental justification has also been presented\cite{2}.
Furthermore, a dynamical system analysis showed that a smooth
transition from inflationary to a radiation phase is attained
for many values of the friction parameter, thereby showing
that the warm inflation scenario may be a workable variant
to standard inflation.
During the warm iflationary era, vacuum fluctuations on scales smaller than
the size of the horizon are magnified into classical perturbations on
scales bigger than the Hubble radius. The classical perturbations can lead
to an effective curvature of spacetime and energy density perturbations
\cite{3}.

In an alternative formalism for warm inflation, I studied a model where
the mean temperature and the amplitude of the temperature's fluctuations
decreases with time for a rapid power - law expanding universe. This is
the most significative difference with the Berera's formalism in which
the warm inflation expansion is isothermal\cite{4,5,6}.
During the warm inflationary expansion, the kinetic energy density
$\rho_{kin}$ is smaller with respect to the vacuum energy
\begin{displaymath}
\rho(\varphi) \sim \rho_m \sim V(\varphi)  \gg \rho_{kin}.
\end{displaymath}
The kinetic energy density is given by
\begin{displaymath}
\rho_{kin} = \rho_r (\varphi) + \frac{\dot\varphi^2}{2},
\end{displaymath}
where
\begin{equation}\label{1}
\rho_r(\varphi) = \frac{\tau(\varphi)}{8 H(\varphi)} \dot\varphi^2.
\end{equation}
Here, the dot denotes the derivative with respect to the time.
Furthermore, $\tau(\varphi)$ and $H(\varphi)$ are the friction
and Hubble parameters. The eq. (\ref{1}) comes from the
assumption that the radiation energy density remains constant
during the inflationary era ($\dot\rho_r \simeq 0$).

The Lagrangian that describes the warm infaltion scenario
is
\begin{equation}
{\cal L}(\varphi, \varphi_{,\mu}) = - \sqrt{-g} \left[
\frac{R}{16 \pi} + \frac{1}{2} g^{\mu\nu}
\varphi_{,\mu} \varphi_{,\nu} + V(\varphi)\right]
+ {\cal L}_{int},
\end{equation}
where $R$ is the scalar curvature, $g^{\mu\nu}$ gives the metric
tensor and $g$ is the metric. The Lagrangian ${\cal L}_{int}$ takes
into account that the particles in the thermal bath interact with
the scalar field $\varphi$.
In principle, a permanent or temporary coupling of the scalar
field $\varphi$ with others fields might also lead to dissipative
processes producing entropy at different eras of the cosmic
evolution. It is expected that progress in nonequilibrium statistics
of quantum fields will provide the necessary theoretical framework
for discussing dissipation in more general cases\cite{10}.

The semiclassical Friedmann equation is
\begin{equation}
H^2(\varphi) = \frac{8\pi}{3 M^2_p} \left<E \left|
\rho_m(\varphi) + \rho_r(\varphi) \right|E\right>,
\end{equation}
where $M_p = 1.2 \times 10^{19}$ GeV is the Planckian mass.

Now I consider the semiclassical expansion for the inflaton field
$\varphi$
\begin{equation}\label{sem}
\varphi(\vec x,t) = \phi_c(t) + \alpha(t) \phi(\vec x,t).
\end{equation}
Here, $\phi_c(t) = <E|\varphi|E>$, $<E|\phi|E> = <E|\dot\phi|E> = 0$,
and $|E>$ is an arbitrary state. Furthermore, $\alpha(t)$ is a dimensionless
time - dependent function that characterize the gravitational coupling
between the fluctuations of the matter field and the fields
in the thermal bath. A lot of work can be done on phenomenological
grounds, as, for instance, by applying nonequilibrium thermodynamic
techniques to the problem or even studying particular models with dissipation.
An example of this latter case is the warm inflationary
picture recently proposed\cite{1}.

The aim of this work is the study of the power spectrum in warm inflation
with the semiclassical expansion (\ref{sem}),
taking into account the COBE data coarse - grained field introduced
in a previous work\cite{8}. This topic was studied in\cite{8} but with
the semiclassical expansion $\varphi = \phi_c + \phi$.
In this work I incorporate
in the formalism the
backreaction of the metric for the study of the effective curvature for the
now observable universe, when the fluctuations are coupled with the thermal
bath.

\section{Dynamics of the inflaton}

\subsection{Dynamics of the classical field}

The dynamics for the classical field in warm inflation was obtained in
previous works\cite{4,5,6}. The equation of motion for $\phi_c$ is
\begin{equation}\label{motio}
\ddot\phi_c + \left[3 H_c +\tau_c\right] \dot\phi_c + V'(\phi_c) =0,
\end{equation}
where $H_c \equiv H(\phi_c) + {\dot a \over a}$, and $\tau_c
\equiv \tau(\phi_c)$ and $V'(\phi_c) \equiv \left.{d V(\varphi) \over
d \varphi}\right|_{\phi_c}$.
The term $\tau_c \dot\phi_c$ in eq. (\ref{motio}) shows as the
scalar field evolves with the time in a damped regime generating
an expansion which depends on the mean temperature $T_r$ of the thermal
bath. As a consequence, the subsequent reheating
mechanism is not needed and thermal fluctuations produce the
primordial spectrum of density perturbations\cite{5,6}.
Furthermore, $\dot\phi_c = - {M^2_p \over 4\pi}
H'_c \left(1+ {\tau_c \over 3 H_c}\right)^{-1}$ and the classical
effective potential is
\begin{eqnarray}
V(\phi_c) &=& \frac{M^2_p}{8\pi} \left[ H^2_c - \frac{M^2_p}{12 \pi}
\left(H'_c\right)^2 \left( 1+ \frac{\tau_c}{4 H_c}\right) \right.\nonumber \\
&\times & \left.\left(1+ \frac{\tau_c}{3 H_c}\right)^{-2}\right].
\end{eqnarray}
The radiation energy density of the background is
\begin{equation}\label{radia}
\rho_r\left[\phi_c(t)\right] \simeq \frac{\tau_c}{8 H_c}
\left(H'_c\right)^2
\left(\frac{M^2_p}{4\pi}\right)^2 \left(1+\frac{\tau_c}{3 H_c}\right)^{-2},
\end{equation}
and the temperature of this background is
\begin{equation}
T_r \propto \rho^{1/4}\left[ \phi_c(t)\right].
\end{equation}
Note that the temperature depends on time. In the warm inflation
model here studied, I will suppose that it decreases with time,
in agreement one expects in an expanding universe.
The temporal evolution of the background temperature depends
on the particular model which one considers. For example, in a power - law
expanding universe $T_r \sim t^{-1/2}$\cite{6}.

\subsection{First order $\phi$ - fluctuations}

In this section I will study the first  order $\phi$ - fluctuations
for the matter field $\varphi$, on a globally flat Friedmann -Robertson -
Walker (FRW) metric
\begin{equation}
ds^2 = - dt^2 + a^2 d\vec x^2,
\end{equation}
which describes a globally isotropic and homogeneous spacetime. The
equation of motion for the quantum perturbations $\phi$, is
\begin{eqnarray}
&& \ddot\phi + \left[ 2 \frac{\dot\alpha}{\alpha} +
(3 H_c+ \tau_c) \right] \dot\phi -\frac{1}{a^2} \nabla^2\phi \nonumber \\
& + & \left[ (3 H_c +\tau_c) \frac{\dot\alpha}{\alpha} +
V''(\phi_c) \right] \phi = 0.\label{mot}
\end{eqnarray}
The function $\alpha(t)$ depends on time.
I consider $\alpha(t) = F[T_r(t)/M]$, where $T_r(t)$ is the temperature
of the background and $M \simeq 10^{15}$ GeV is the GUT mass.
The structure of the equation (\ref{mot}) can be simplified
by means of the map $\chi = e^{3/2 \int \left
(H_c + \tau_c/3 +{2\dot\alpha\over 3 \alpha}\right)dt} \  \phi$
\begin{equation}
\ddot\chi - \frac{1}{a^2} \nabla^2 \chi - \mu^2(t) \chi =0,
\end{equation}
where $\mu^2(t)= {k^2_o \over a^2}$ is the time dependent
parameter of mass and  $k_o(t)$ is the time dependent
wave number which separates the long wavelength ($k \ll k_o$)
and the short wave ($k \gg k_o$) sectors.

The square time dependent parameter of mass is
\begin{equation}
\mu^2(t)  =  \frac{9}{4} \left( H_c + \frac{\tau_c}{3}
\right)^2 - V''(\phi_c)
+\frac{3}{2} \left(\dot H_c + \frac{\dot\tau_c}{3} \right).
\end{equation}
Note that $\mu(t)$ do not depends on the function $\alpha(t)$.

\subsection{Second order $\phi$ - fluctuations and Backreaction}

Making a second order $\phi$ - fluctuations expansion for $\varphi$, one
obtains the following semiclassical Friedmann equation
\begin{equation}
H^2_c + \frac{K}{a^2} = \frac{8\pi}{3 M^2_p} \left<
E \left| \rho_m + \rho_r \right| E \right>,
\end{equation}
where $K$ is an effective curvature produced by the backreaction of the
metric with the fluctuations of the scalar field. This curvature is
given by
\begin{eqnarray}
\frac{K}{a^2} & = & \frac{8\pi}{3 M^2_p} \left[\left(
1+ \frac{\tau_c}{8 H_c}\right) \left( \frac{\dot\alpha^2}{2}
\left<\phi^2\right>
+ \frac{\alpha^2}{2} \left< \dot\phi^2 \right>\right.\right. \nonumber \\
&+&\left.\left. \alpha \dot\alpha \left< \phi \dot\phi\right> \right) +
\frac{\alpha^2}{a^2}  \left< \left(\vec\nabla \phi\right)^2 \right>
+ \frac{V''}{2} \alpha^2 \left< \phi^2 \right> \right].
\end{eqnarray}
Note that $K$ depends on the temporal evolution of $\alpha(t)$
as well as the expectation values for $\phi^2$, $\phi\dot\phi$,
$\dot\phi^2$ and $\left(\vec\nabla \phi\right)^2$. If $\alpha(t)$
is a function of the temperature, $\alpha(t) = F[T_r(t)/M]$, the
instantaneous comoving temperature will be very important during
the warm inflationary regime.

To study the backreaction of the metric with the fluctuations
$\phi$, I introduce the metric
\begin{equation}
ds^2 = -dt^2 + a^2 \left[ 1+ h(\vec x,t)\right] d\vec x^2,
\end{equation}
where $h(\vec x,t)$ represents the fluctuations of the metric
produced by the $\phi$ - fluctuations.
Making the following expansion for $H(\varphi)$
\begin{equation}
H[\varphi(\vec x,t)] \simeq H_c[\phi_c(t)]
+ H'[\phi_c(t)] \  \alpha(t) \  \phi(\vec x,t),
\end{equation}
one obtains the following expression for $h(\vec x,t)$\cite{3}
\begin{equation}
h(\vec x,t) \simeq 2 {\large\int}^{t} dt' \  \alpha(t') \phi(\vec x,t')
H'[\phi_c(t')],
\end{equation}
and the effective curvature can be represented by
\begin{eqnarray}
\frac{K}{a^2} &=& \left< E \left| \left(\frac{\dot h(\vec x,t)}{2}\right)^2
\right|E \right> \nonumber \\
& = & \left< E \left| \left[\alpha(t) \phi(\vec x,t) H'(\phi_c)\right]^2
\right|E \right>. \label{curv}
\end{eqnarray}
This expression shows that the temporal evolution of the effective curvature
arises from the matter field fluctuations $\phi(\vec x,t)$ and the
temperature of the thermal bath, due to the fact I am considering
that $\alpha(t)$ is a function of the temperature of this bath.
In order to study the evolution of the fluctuations on the infrared
(long wavelength) sector, firstly one can write the fields $\chi$ and
$h$ as two Fourier expanded fields
\begin{eqnarray}
\chi(\vec x,t) &=& \frac{1}{(2\pi)^{3/2}} {\large\int} d^3k \  \left[
a_k \chi_k(\vec x,t) + a^{\dagger}_k \chi^{*}_k(\vec x,t) \right], \\
h(\vec x,t) &=& \frac{1}{(2\pi)^{3/2}} {\large\int} d^3k \  \left[
a_k h_k(\vec x,t) + a^{\dagger}_k h^{*}_k(\vec x,t) \right],
\end{eqnarray}
where $ \chi_k(\vec x,t) =e^{{\rm i} \vec k . \vec x} \xi_k(t)$ and
$ h_k(\vec x,t) =e^{{\rm i} \vec k . \vec x} \tilde\xi_k(t)$.
Here, $\tilde\xi_k(t) = 2 \int^{t}
dt' \  \alpha(t') \ H'[\phi_c(t')] \xi_k(t')$ and
the asterisk denote the complex conjugate. The operators $a_k$ and
$a^{\dagger}_k$ are the well known annihilation and creation operators,
which satisface $[a_k,a^{\dagger}_{k'}]=
{\rm i} \delta^{(3)}(k-k')$ and $[a^{\dagger}_k,a^{\dagger}_{k'}]=
[a_k,a_{k'}]=0$.
The commutation relations for the fields $\chi$ and $h$ are
\begin{eqnarray}
&&\left[\chi(\vec x,t), \dot\chi(\vec x',t)\right] = \nonumber \\
&& \frac{1}{(2\pi)^3} {\large\int} d^3k \left(\xi_k \dot\xi^*_k -
\dot\xi_k \xi^*_k\right) e^{-{\rm i} \vec k.(\vec x-\vec x')},
\label{comut}\\
&& \left[h(\vec x,t), \dot h(\vec x',t)\right] = \nonumber \\
&& \frac{1}{(2\pi)^3} {\large\int} d^3k
\left(\tilde\xi_k \dot{\tilde\xi}^*_k -
\dot{\tilde\xi}_k \tilde\xi^*_k\right)
e^{-{\rm i} \vec k.(\vec x-\vec x')}.
\end{eqnarray}
To obtain $\left[\chi(\vec x,t), \dot\chi(\vec x',t)\right]={\rm i}
\delta^{(3)}(\vec x - \vec x')$ in eq. (\ref{comut}), one
requires that $\left(\xi_k \dot\xi^*_k -\dot\xi_k \xi^*_k\right) =
{\rm i}$.
\section{Data COBE coarse - grained fields and Stochastic representation}

The data COBE coarse - grained matter field $\chi_{Ccg}$ were introduced
in a previous work\cite{5}
\begin{equation}\label{co1}
\chi_{Ccg} = \frac{1}{(2\pi)^{3/2}} {\large\int} d^3k \  G(k,t)
\left[a_k \chi_k + a^{\dagger}_k \chi^*_k\right].
\end{equation}
Now we can introduce the data COBE coarse - grained field
$h_{Ccg}$ for the fluctuations of the metric
\begin{equation}\label{co2}
h_{Ccg} = \frac{1}{(2\pi)^{3/2}} {\large\int} d^3k \  G(k,t)
\left[a_k h_k + a^{\dagger}_k h^*_k\right].
\end{equation}
In eqs. (\ref{co1}) and (\ref{co2}) the suppression factor $G(k,t)$
is given by\cite{8}
\begin{equation}\label{sup}
G(k,t) = \sqrt{\frac{1}{1+ \left(\frac{k_o(t)}{k}\right)^N}},
\end{equation}
with $N=m-n$. Causality places a strict constraint on the
suppression index: $N \ge 4-n$.
A suppression factor like (\ref{sup}) also was founded
in a model with cosmic strings plus cold or hot
dark matter\cite{*}
Furthermore, the square fluctuations
for the data COBE coarse - grained matter field is
\begin{eqnarray}
\left<E\left| \chi^2_{Ccg}\right|E\right> &=&
{\large\int}^{\infty}_{0} \frac{dk}{k} \  {\cal P}_{\chi_{Ccg}}(k) \nonumber \\
& = & \frac{1}{2\pi^2} {\large\int}^{k_o(t)}_{0} dk \  k^2
\left|\xi_k(t)\right|^2 \  G^2(k,t),
\end{eqnarray}
where the power spectrum ${\cal P}_{\chi_{Ccg}}(k)$ when
the horizon exit is\cite{9}
\begin{equation}\label{po}
{\cal P}_{\chi_{Ccg}}(k) = A(t_*) \left(\frac{k}{k_o(t_*)}\right)^n
f(k).
\end{equation}
Here, $t_*$ denotes the time when the horizon entry, for which $k_o(t_*)
\simeq \pi H_o$ in comoving scale.
The parameters in eq. (\ref{po}) are the amplitude $A(t_*)$
on time $t_*$, the spectral index $n$, the suppression wavenumber
$k_o$ and the suppression index $m$.

The stochastic equation for $\chi_{Ccg}$ is\cite{8}
\begin{equation}\label{sto}
\ddot\chi_{Ccg} - \left(\frac{k_o(t)}{a(t)}\right)^2 \chi_{Ccg} =
\frac{N}{k_o(t)} \left[\xi_1+\xi_2\right],
\end{equation}
where
\begin{eqnarray}
\xi_1(\vec x,t) &=& - \frac{\dot k_o k_o^N}{(2\pi)^{3/2}}
{\large\int} d^3 k \  k^{-N} \  G^3(k,t) \nonumber \\
&\times & \left[a_k \dot\chi_k + a^{\dagger}_k \dot\chi^*_k\right],
\label{no1} \\
\xi_2(\vec x,t) &=& - \frac{\dot k_o k_o^N}{(2\pi)^{3/2}}
{\large\int} d^3 k \  k^{-N} \  G^5(k,t) \nonumber \\
&\times& \left[\left(\frac{k_o}{k}\right)^N
\left(3 \dot k^2_o - 2 k_o \ddot k_o\right) + 2 \dot k^2_o (1-N)
- 2 k_o \ddot k_o \right]  \nonumber \\
&\times & \left[a_k \chi_k + a^{\dagger}_k \chi^*_k\right].\label{no2}
\end{eqnarray}
The noises (\ref{no1}) and (\ref{no2}) arise from the increasing
number of degrees of freedom of the ifrared
sector from the short-wavelength sector. For the
special case considered in eq. (\ref{sup}), $\xi_1$ is a
colored noise, while $\xi_2$ gives non - local dissipation.

Since $h \simeq 2 \int^t dt' \alpha(t')
H'(t') \phi(\vec x,t)$, one can rewrite it as $h\simeq 2 \int^t dt'
\tilde\alpha(t') H'(t') \chi(\vec x,t)$, where $\tilde\alpha(t) =
e^{-3/2\int dt(H_c+\tau_c/3+{2 \dot\alpha \over 3 \alpha})} \alpha(t)$.
With this representation for $h$, the data COBE coarse - grained
metric field $h_{Ccg}$ becomes
\begin{equation}\label{hg}
h_{Ccg}(\vec x,t)
\simeq 2 {\large\int}^t dt' \  \tilde\alpha(t') \  H'(t') \  \chi_{Ccg}
(\vec x,t').
\end{equation}
Replacing (\ref{hg}) in eq. (\ref{sto}), one obtains the
following stochastic equation for $h_{Ccg}$
\begin{eqnarray}
&& \frac{\partial^3 h_{Ccg}}{\partial t^3}
+ 2 \frac{\dot g(t)}{g(t)} \frac{\partial^2 h_{Ccg}}{\partial t^2}-
\left(\frac{k_o}{a}\right)^2 \frac{\partial h_{Ccg}}{\partial t} \nonumber \\
&+& \frac{\ddot g(t)}{g(t)} h_{Ccg}
=\frac{N}{g(t) k_o(t)} \left[ \xi_1(\vec x,t) + \xi_2(\vec x,t)\right],
\label{stochas}
\end{eqnarray}
where $g(t) = \left[2 \tilde\alpha(t) H'(t)\right]^{-1}$.
The square fluctuations for the field $\phi_{Ccg}$ is
\begin{eqnarray}
\left<E \left| \phi^2_{Ccg}\right|E\right> & = &
\frac{e^{-3 \int \left(H_c+\frac{\tau_c}{3}
+\frac{2 \dot\alpha}{3\alpha}\right)dt}}{2 \pi^2} \nonumber \\
&\times & {\large\int}^{k_o}_0 dk \  k^2 \left|\xi^2_k(t)\right| G^2(k,t).
\label{fluct}
\end{eqnarray}
Thus, the effective curvature $K/a^2$ for the now observable
universe is [see eq. (\ref{curv})]
\begin{equation}
\left.\frac{K}{a^2}\right|_{\rm COBE} =
\left[\alpha(t) \  H'[\phi_c(t)]\right]^2
\left<E \left| \phi^2_{Ccg}\right|E\right>.
\end{equation}
Hence, the power spectrum for $\phi_{Ccg}$ and $h_{Ccg}$ when the
horizon entry, are
\begin{eqnarray}
{\cal P}_{\phi_{Ccg}}(k) & = & B(t_*) \left(\frac{k}{k_o(t_*)}\right)^n
f(k), \\
{\cal P}_{h_{Ccg}}(k) & = & C(t_*) \left(\frac{k}{k_o(t_*)}\right)^n
f(k).
\end{eqnarray}
Here, $B(t_*)$ and $C(t_*)$ are the amplitude such that
\begin{eqnarray}
B(t_*) & = & A(t_*) \  e^{-3 \int^{t_*} \left(H_c+\frac{\tau_c}{3}
+\frac{2 \dot\alpha}{3\alpha}\right)dt}, \label{al}\\
C(t_*) &=& \left.\left[\alpha(t) \  H'[\phi_c(t)]\right]^2 \right|_{t=t_*}
B(t_*). \label{be}
\end{eqnarray}
Due to $\left|\delta_k\right|^2 = {\cal P}_{\phi_{Ccg}}(k)$\cite{8},
the spectral density become $|\delta_k| = k^n f(k)$. The standard
choice $n=1$ and $f(k)$ as constant,
was invoked by Harrison\cite{ha} and Zel'dovich\cite{ze}
on the grounds that it is scale invariant at the epoch of the
horizon entry. The constraint $|n-1| < 0.3$ was obtained from
the data COBE spectrum\cite{*}. Note that both $B(t_*)$ and $C(t_*)$
depends on the temperature
of the background when the horizon entry. This is a very important
characteristic that becomes from this formulation, once one
consider $\varphi = \phi_c + \alpha(t) \phi$ and $H(\varphi)=
H_c + \alpha(t) H' \phi$ as semiclassical
expansions for $\varphi$ and $H(\varphi)$.
From eq. (\ref{be}) one obtains
\begin{equation}
\frac{C(t_*)}{B(t_*)} =
\left.\left[\alpha(t) \  H'[\phi_c(t)]\right]^2 \right|_{t=t_*}.
\end{equation}
Taking $\rho_r(t_*)={\pi^2 \over 30} N[T_r(t_*)] \  T^4_r(t_*)$,
where $N[T_r(t_*)]$ is the number of relativistic degrees of freedom
at temperature $T_r(t_*)$ and replacing $(H'_c)^2$ in
eq. (\ref{radia}), one obtains (for $N[T_r(t_*)] \simeq 10^3$, $\alpha(t_*)=
\left({T_r(t_*) \over M}\right)^{\beta}$ --- $\beta \ge 0$ ---
and $M \simeq 10^{-4} M_p$)
\begin{equation}
\frac{C(t_*)}{B(t_*)} = \frac{64}{45} \pi^4 \  10^{3+8\beta}
\frac{(3H_c+\tau_c)^2}{3 H_c\tau_c}
\left(\frac{T_r(t_*)}{M_p}\right)^{
2(\beta+2)}.
\end{equation}
For the case $\tau_c(t_*) \simeq H_c(t_*)$, one obtains the expression
\begin{equation}
\frac{C(t_*)}{B(t_*)} \simeq
10^{(6+8\beta)}
\left(\frac{T_r(t_*)}{M_p}\right)^{
2(\beta+2)}.
\end{equation}
For example, for $\beta=1$ one obtains $T_r(t_*) = 10^{-5} M_p$ for
${C(t_*) \over B(t_*)} \simeq 10^{-15}$.
This implies that the amplitude for the fluctuations of the metric
when the horizon entry,
should be very small for the expected values of temperature.

\section{Final Remarks}

To summarize, in this letter I considered a model for warm inflation
where the fluctuations of the scalar field are coupled with the
thermal bath. This coupling, depends on the temperature of the background
which is a function of the temperature. The temperature
decreases with the time,
as well as the Hubble parameter.

By means of the COBE data coarse - grained field for the fluctuations of the
scalar field I characterize these fluctuations on the scale of the now
observable universe. Once on knows the stochastic equation for
the field $\chi_{Ccg}$, it is possible to obtain the stochastic
equation for $h_{Ccg}$ [see eq. (\ref{stochas})]. The square fluctuations
for $\phi_{Ccg}$ and $h_{Ccg}$ give the spectral density $\delta_k$ for both,
$\phi_{Ccg}$ and $h_{Ccg}$. The spectral density
$\delta_k$ depends on the function $G^2(k,t)$, the modes $\xi_k$
and the index $n$.
Finally,
I find that the
amplitude for the fluctuations of the metric
when the horizon entry,
should be very small for the expected values of temperature.

\end{document}